# Spin frustration effects in an odd-member antiferromagnetic ring and the magnetic Möbius strip

Olivier Cador,[a] Dante Gatteschi,[a] Roberta Sessoli,[a]* Anne-Laure Barra,[b] Grigore A. Timco[c] and Richard E. P. Winpenny[c]

[a]*LAboratory of Molecular Magnetism, Department of Chemistry & INSTM, Università degli Studi di Firenze, Via Lastruccia n. 3, 50019 Sesto Fiorentino, Italy*
[b]*Laboratoire des Champs Magnetiques Intenses-CNRS, F-38042 Grenoble Cede 9, France*
[c]*Department of Chemistry, The University of Manchester, Oxford Road, Manchester, M13 9PL, UK.*

**Abstract** The magnetic properties of the first odd-member antiferromagnetic ring comprising eight chromium(III) ions, S=3/2 spins, and one nickel(II) ion, S=1 spin, are investigated. The ring possesses an even number of unpaired electrons and a S=0 ground state but, due to competing AF interactions, the first excited spin states are close in energy. The spin frustrated ring is visualized by a Möbius strip. The "knot" of the strip represents the region of the ring where the AF interactions are more frustrated. In the particular case of this bimetallic ring electron paramagnetic resonance (EPR) has unambiguously shown that the frustration is delocalized on the chromium chain, while the antiparallel alignment is more rigid at the nickel site.

## 1. Introduction

One aim of molecular chemistry is to produce, by design, complex molecules that display new phenomena[1]. These phenomena might allow testing of fundamental theories or might have potential technological applications. For example, the successful use of "single molecule magnets", to test fundamental theories, such as quantum tunneling of the magnetization has prompted a rapid development of molecular nanomagnetism[2-4]. However the concentration of research on these molecules has perhaps distracted the scientists from looking for other properties in magnetic molecules, and from targeting metal arrays that might allow further physical phenomena to be examined in detail. One of the most interesting phenomena is spin frustration [5], which occurs when all the interactions between spin pairs cannot simultaneously have their optimal value. The simplest systems where this could be observed are an odd-member ring of antiferromagnetically coupled spins.

Even member rings of antiferromagnetic spins are quite common for transition metal ions, like Fe(III)[6-10], Cr(III)[11,12], Cu(II)[13], or Mn(III)[14]. Antiferromagnetic iron rings with N= 6, 8, 10, 12, and 18 have been reported and their properties analyzed in some detail. Beyond being of interest as models for low dimensional magnets, antiferromagnetic rings are attracting interest for the hypothesis that they may provide good opportunities for observing quantum coherence in the fluctuation of the Néel vector[15].

On the contrary odd member rings, excluding the case of three spins, are practically unprecedented. A possible explanation could be that the most relaxed structure for rings has the bridging ligands alternately located above and below the plane defined by the metal ions. An odd number of metal sites introduces a sort of incommensurability that destabilizes the structure. Additional interactions, like the hydrogen bond with a host cation inside the ring, can overcome this problem, as recently shown in odd derivatives of Cr(III) rings[16].

As the interactions with the host cation plays a crucial role in stabilizing odd structures the ring must be negatively charged and this can be achieved only in hetero-metallic rings, for instance where a M(II) ion substitutes a Cr(III) one. This is the case of the recently reported $Cr_8Ni$ compound of formula $[(C_6H_{11})_2NH_2][Cr_8NiF_9(O_2CCMe_3)_{18}]$[16], whose structure is reported in Figure 1.

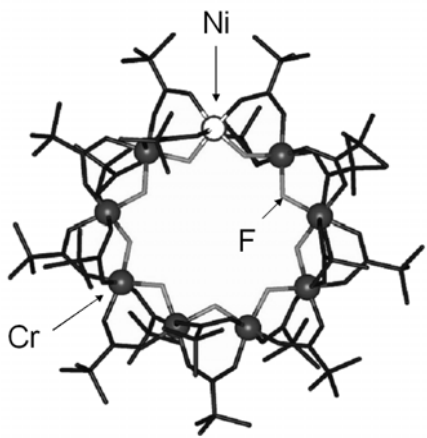

**Figure 1.** Structure of Cr8Ni ring. The metal ions are bridged by one fluoride ion and two $(CH_3)_3COO^-$ ions.

This is the first antiferromagnetic ring with an odd number of spins whose magnetic properties, and in particular the spin frustration effects, have been recently communicated [16] and will be discussed in more details in the following sections.

## 2. Magnetic Properties of $Cr_8Ni$.

The temperature dependence of the molar susceptibility of a polycrystalline sample of Cr8Ni is presented in Figure 2. The magnetic susceptibility goes through a round maximum at ca. 25 K, as often observed in AF Cr(III) rings. Below this temperature the susceptibility goes through a large minimum and increases again to pass through a second maximum around 2.0 K. The presence of two maxima is unprecedented in molecular systems and suggests that the ground state is non-magnetic but very close in energy to excited magnetic states.

The field dependence of the magnetization has been measured up to 120 kOe and the results are plotted in Figure 3. The magnetization reaches a plateau at around 2 $\mu_B$ and 80 kOe and then increases again. Interestingly the derivative of the magnetization curve, also reported in Figure 3, shows a first maximum at ca. 30 kOe and a second abrupt increase above 90 kOe.

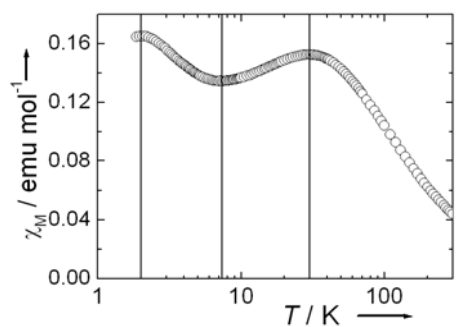

**Figure 2.** Variation of $\chi_m$ with temperature for $Cr_8Ni$.

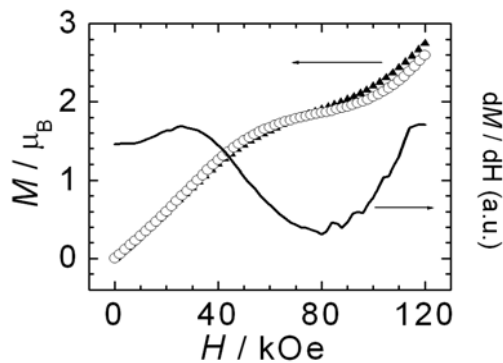

**Figure 3**. Field dependence of the magnetization for $Cr_8Ni$ measured at 1.6 K (○) and 2.0 K (▲). The solid line is the numerical derivative of the curve at 1.6 K.

The observed behavior can be qualitatively rationalized assuming that the ground state is not magnetic and that the first excited state has S=1 and is therefore so close in energy that the applied magnetic field is able to stabilize it, thus inducing a cross-over of the states. The cross-over field roughly coincides with the maximum in the dM/dH curve, ca 30 kOe. The further increase in the derivative above 90 kOe indicates that another spin state with S> 1 starts to be populated as an effect of the applied magnetic field [6,17,18].

In order to get more information on the nature of the lowest lying spin states electron paramagnetic resonance (EPR) spectra were recorded on a laboratory made high field-high frequency spectrometer[19]. The use of a high field has been chosen to enhance the accuracy in the determination of the g factor.

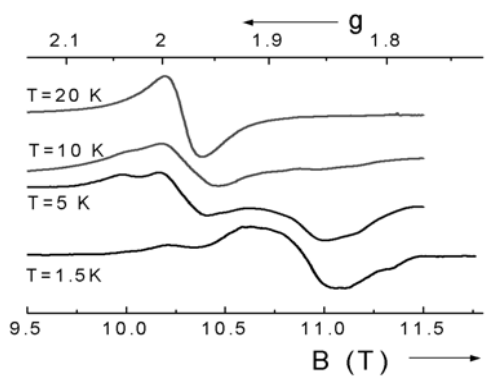

**Figure 4.** Polycrystalline HF-EPR spectra recorded at 285 GHz and four different temperatures with a laboratory-made spectrometer based on a Gunn-diode as far-infrared frequency source.

In Figure 4 are reported the spectra of a polycrystalline powder recorded at different temperatures with an exciting radiation of 285 GHz. At 20 K the spectrum consists of a symmetric line centered around g=1.98, but decreasing the temperature a significant g-shift is observed.

At 1.5 K the main absorption is around g=1.86 while the line becomes more complex.

**3. Spin frustration effect in Cr8Ni**

The investigated ring contains an odd number of spins but they are not all identical to each other, as would be expected in the ideal model of spin frustration. Nevertheless not all the AF magnetic interactions of the rings can be simultaneously satisfied and spin frustration is therefore present in Cr8Ni.

We can assume, in a first approximation, that only two different exchange interactions are active in the ring, namely that related to Cr-Cr pairs and that to Cr-Ni pairs. Only nearest neighbor interactions are considered as schematized in Figure 5.

In the parent octanuclear Cr(III) compound $[Cr_8F_8(O_2CCMe_3)_{16}]$, Cr8, the S=3/2 are antiferromagnetically coupled and the exchange coupling constant $J/k_B$ is ca. 17 K[12]. The ground state is of course S=0 and the susceptibility goes through a round maximum at ca. 40 K. The curve of Figure 2 is reminiscent at high temperature of that observed for Cr8, suggesting that a similar antiferromagnetic interaction is active between Cr spins. The inclusion of the S=1 spin of Ni(II) in the ring maintains the number of unpaired electrons even, thus S=0 can again be the ground state, as suggested from the experimental data of Figure 2. However, the ground state strongly depends on the nature and strength of the Cr-Ni interaction. If no interaction is active the low temperature magnetic behavior is simply that of the isolated S=1 of the nickel spin, obeying the Curie law and in disagreement with the observed behavior.

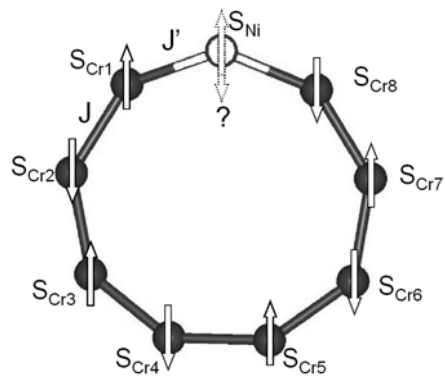

**Figure 5.** Labeling scheme and spin structure for Cr8Ni.

Interestingly spin frustration appears as soon as a magnetic interaction is switched on between Ni and Cr, independent of the sign of this interaction. In fact the edge spins of the Cr chains are antiparallel to each other but the interaction with the spin of Ni tends to orient them parallel to each other, either antiparallel or parallel to the Ni spin, depending if $J_{CrNi}$ is antiferromagnetic or ferromagnetic, respectively.

Visualization of spin frustration is always difficult. We have found that the spin structure of the Cr8Ni ring has some analogy with the Möbius strip. The Möbius strip contains a knot that in our visualization represents the frustrated interaction. Regions where the strip is vertical correspond to regions in the ring where the AF interactions are satisfied, while the horizontal parts represent the frustrated interactions. The knot, and thus the frustrated interactions, can be localized on the Ni site, or can be delocalized on the Cr chain. These two possibilities are schematized in Figure 6, where the Ni site is evidenced in white.

hamiltonian:

$$H = J'(\mathbf{S}_{Cr_1} \cdot \mathbf{S}_{Ni} + \mathbf{S}_{Cr_8} \cdot \mathbf{S}_{Ni}) + J\sum_{i=1}^{7}\mathbf{S}_{Cr_i} \cdot \mathbf{S}_{Cr_{i+1}}$$
(1)

where the labeling is that of Figure 5.

In order to reduce the size of the calculation a method based on the Irreducible Tensor Operators has been employed[20]. The matrix is thus factorized in blocks defined by the total spin value S, which ranges from 0 to 13.

We have first tried to reproduce the very peculiar χ vs. T curve of Figure 2. In Figure 7 we report the calculated values assuming J=16 K, similar to what has been observed in Cr8 ring, and varying J' between 10 and 70 K. A ferromagnetic J' was also considered but this resulted in a ground S=1 state, as shown in Figure 8, where the energies of the first S=0 and S=2 and the first three S=1 states as a function of J' are reported.

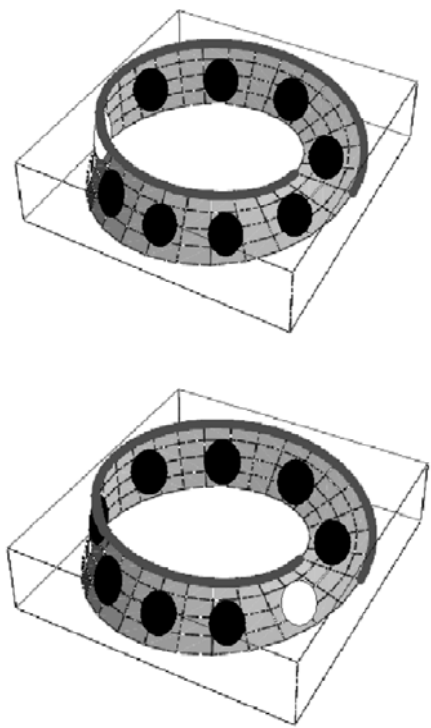

**Figure 6.** Representation of the spin frustration in Cr8Ni as a Möbius strip, with the white circle as the Ni site, and black circles as Cr: (top) with J'<< J and the "knot" on the Nickel site; (bottom) with J'>>J and the "knot" on the chromium spins chain. The "knot" is the point at which the upper line is discontinuous.

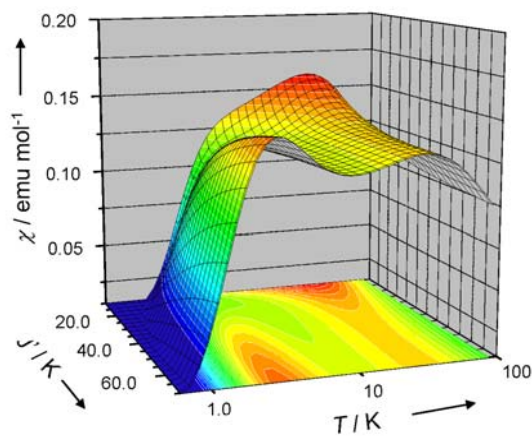

**Figure 7.** Temperature dependence (in log scale) of the calculated magnetic susceptibility using the Hamiltonian (1) with J fixed to 16 K and varying J' in the range 10-70 K. An average g=2 has been assumed

The delocalization of spin frustration in the ring depends on the relative strength of the two interactions. If $J_{CrNi}$ is stronger than $J_{CrCr}$, spin frustration is mainly delocalized on the Cr chain, and vice versa.

In order to distinguish between the two situations it is necessary to have a deeper insight into the energy spectrum of the lowest lying states, as shown in the following section.

## 4. Quantitative analysis of the magnetic properties

In order to rationalize the observed behavior we have calculated the energy spectrum and the magnetic susceptibility of Cr8Ni by using the following spin

It is evident from Figure 8 that the ground state changes as a function of J': it is S=2 for J'<-2.9 K, S=1 for –2.9<J'<1.5 K, and S=0 above this value. In the range 3-6 K we observe several level crossings and the first S=1 state changes its nature in this region.

From Figure 7 we evidence that the two maxima in the susceptibility experimentally observed are only reproduced if J' is larger than 40 K and therefore significantly stronger than J. A reasonable agreement with the data of Figure 2 is obtained with J'=70 K even if a true fitting procedure was not attempted as explained in the following.

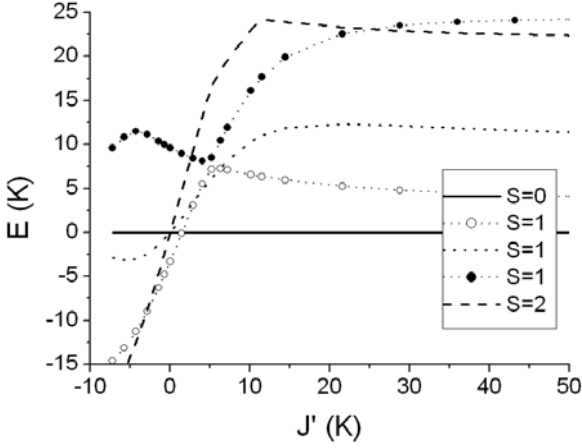

**Figure 8.** Energy of the first S=0, the first three S=1 and the first S=2 spin states as a function of J' calculated using J=16 K in eq (1). E(S=0) has been chosen as the zero. The two S=1 states that show an avoided level crossings for J'≈5 K are evidenced with empty and filled circles. Other states are present in the reported energy window but not shown for the sake of clarity.

The calculated energy spectrum reveals that with these parameters the ground S=0 state is only 3.7 K below the first S=1, while the second excited state, again S=1, is at ≈ 10.7 K from the ground state. The first $S = 2$ is calculated to occur at ≈ 22 K above the ground state. This energy scheme is not significantly varied if J' is reduced by a factor of two (see Figure 8), as usual in the presence of spin frustration.

Such an energy scheme is in agreement with the magnetization curve of Figure 3, and in particular with the observed maximum at ca 30 kOe, because at this field $g\mu_B H \approx \Delta E(S=1,S=0)$ [21].

The estimated value of J' suggests that the frustrated interactions are those within the Cr spins chain rather than those involving the Ni spin. The Möbius strip in the upper part of Figure 6 therefore describes the system.

The type of frustration reflects also on the wave functions of the first spin states, and it is therefore interesting to evaluate the eigen-vectors of the low-lying states. In this particular case it is very convenient to use for the basis set the representation where we couple the chromic spins on odd sites to give an intermediate spin $S_{odd}$ and analogously for the spins on the even sites, $S_{even}$. These two intermediate spins are coupled together, to give the total spin for the chromium chain, $S_{TCr}$, and this last one is coupled to the spin of the nickel to give the total spin, $S_T$. There are 2764 different ways of obtaining an $S_T=1$ state,

therefore the wave functions of Hamiltonian (1) are linear combinations:

$$\Psi_k = \sum_{i=1}^{2764} c_i^k \varphi_i \qquad (2)$$

where every $\varphi_i$ is defined by seven intermediate spin quantum numbers as $|S_{1-3},S_{13-5},S_{odd},S_{2-4},S_{24-6},S_{even},S_{TCr},S_T=1\rangle$. However, for the states lowest in energy only few $c_i$'s are significantly different from zero. The composition of the first excited $S_T = 1$ state strongly depends on the J'/J ratio. If J'<<J $\psi_1$ is mainly given by $S_{even} = 6$, $S_{odd} = 6$, with $S_{TCr}=0$ and can be seen as mainly given by the uncorrelated spin of the nickel ion. The frustration, or the "knot", is localized on the nickel site, as shown in the lower part of Figure 6. On the contrary when J'>>J the largest contribution comes from $S_{TCr}=2$ antiferromagnetically coupled with the nickel spin to give $S_T=1$. In this case the antiferromagnetic order is more rigid at the nickel site and the frustration is instead delocalized on the chromium chain, as shown in the upper part of Figure 6.

Usually magnetic measurements do not provide direct information on the wave-function composition of the spin states. On the contrary the g value of the different spin state results from a linear combination of the individual g values

$$g(\Psi_k) = \sum_{n=1}^{9} a_n^k g_n \qquad (3)$$

where $n$ refers to the spin site in the ring and $a_n$ are normalized coefficients that can be obtained by spin projection techniques[22]. The calculation of these coefficients is relatively simple if the wave function $\psi_k$ coincides with one element of the basis set, $\varphi_i$. Otherwise each $a_n^k$ results from a weighted summation of all the $a_n^i$, where the weight is given by the coefficients of eq (2).

If we consider the limiting case where J>>J' then the first $S_T=1$ state is well described by the basis function $|3,9/2,6,3,9/2,6,0,1\rangle$ and $a_1=...=a_8=a_{Cr}=0$ and $a_9=a_{Ni}=1$. The g value of the S=1 state should then be equal to that of nickel: $g_{S=1} = g_{Ni}$.

On the contrary, if J'>>J, the first $S_T=1$ state is well described by $|3,9/2,6,3,9/2,6,2,1\rangle$. Spin frustration is denoted by the fact that $S_{even}$ and $S_{odd}$ are not fully antiparallel to each other, resulting in $S_{Cr}=2$. In this last case by applying the projection techniques we obtain $g_{S=1}=3/2 g_{Cr}-1/2 g_{Ni}$.

If both metal ions are characterized by the same g values no difference can be detected between the two cases described above. However, in this bimetallic ring we have $g_{Ni} \approx 2.2$ that is significantly larger than $g_{Cr} \approx 1.98$. In the two limiting cases mentioned above the g of the first S=1 state is therefore 2.2 and 1.87, respectively.

HF-EPR spectroscopy is for sure one of the most efficient techniques to determine the g values with great accuracy. The spectra shown in Figure 4 suggest that on lowering the temperature the signal shifts to higher field, i.e. lower g values, as expected if the S=1 with g=1.87 becomes, the most populated magnetic state on lowering the temperature. This suggests that J'>>J in agreement with the simulated magnetic susceptibilities.

The experimental susceptibility has not been fitted because a reasonable fitting should take into account different g values of the different spin states. Such a calculation would require taking into account the wave function composition for every state and also calculating the projection coefficients for every element of the basis of every spin value, in total 23548 spin states. This is a formidable task to be included in a minimization procedure.

## 5. Conclusions

The potentiality of molecular synthesis to provide model magnetic systems is evident in the present case, where the first AF magnetic ring with an odd number of spins has been obtained and magnetically characterized. Even if its structure does not coincide with the ideal model of spin frustration, i.e. an odd number of s=1/2 spins, it surely shows the effects of competing interactions, which reveals the presence of excited magnetic states very close in energy to the ground non-magnetic state. Moreover, the presence of two different spins in the ring characterized by a significantly different Landè factor, has allowed to unambiguously determine how frustration is distributed in the ring: if it is localized on the nickel site or rather more distributed on the chromium chain. Often this information can only be retrieved through Inelastic Neutron Scattering experiments, while here more affordable techniques, magnetization measurements and HF-EPR spectroscopy, have provided a detailed rationalization of the magnetic properties.

Spin frustrated rings surely deserve further theoretical and experimental investigation. It would be extremely interesting to investigate the dynamics of the Néel vector when frustration is present[15], as well as the nature of the crossings induced by the field, in particular if these are real crossings or rather avoided crossings due to admixing of spin states[23,24].

## 6. Acknowledgments

We acknowledge the financial support of the EPSRC (UK), of the EC through the EC-TMR Network and "QuEMolNa" (MRTN-CT-2003-504880), of the German DFG (SPP 1137), and of INTAS.